\documentclass[aps,pra,floatfix,superscriptaddress,preprint, showpacs]{revtex4}

\usepackage{graphicx}
\usepackage{graphics}
\usepackage{amssymb}
\usepackage{amsmath}
\usepackage{epsfig}
\usepackage{latexsym}
\usepackage{color}
\usepackage{rotating}
\usepackage{subfigure}
\usepackage{float}

 \newcommand{\p}{\mathrm{P}}

\begin{document}

\title{Classical and Quantum Causal Interventions }

\author{G. J. Milburn \footnote{Author to whom any correspondence should be addressed. Electronic mail: milburn@physics.uq.edu.au} and Sally Shrapnel.}

\affiliation{Centre for Engineered Quantum Systems, School of Mathematics and Physics, The University of Queensland, St Lucia, QLD 4072, Australia}

\begin{abstract}
Characterising causal structure is an activity that is ubiquitous across the sciences. \emph{Causal~models} are representational devices that can be used as oracles for future interventions, to~predict how values of some variables will change in response to interventions on others. Recent~work has generalised concepts from this field to situations involving quantum systems, resulting in a new notion of quantum causal structure. A key concept in both the classical and quantum context is that of an \emph{intervention}. Interventions are the controlled operations required to identify causal structure and ultimately the feature that endows causal models with empirical meaning. Although interventions are a crucial feature of both the classical and quantum causal modelling frameworks, to~date there has been no discussion of their physical basis. In this paper, we consider interventions from a physical perspective and show that, in both the classical and quantum case, they are constrained by the thermodynamics of measurement and feedback in open systems. We~demonstrate that the perfect ``atomic'' or ``surgical'' interventions characterised by Pearl's famous do-calculus are physically impossible, and this is the case for both classical and quantum systems.
\end{abstract}


\maketitle

\section{Introduction}
Causal reasoning is indispensable in science, medicine, economics, and many aspects of every day life. Identifying when one event is the cause of another and how intervening can modify future events are activities that move beyond mere statistical prediction \cite{Spirtes,Pearl,Woodward}.

The task of identifying causal relations directly from empirical data has received much attention in recent decades \cite{Peters}. Economists, computer scientists, statisticians, and philosophers have all contributed to the modern field of causal inference, where the overarching goal is to produce so-called \emph{causal models}, graphical devices that can be used to support causal queries about various phenomena of interest. Much of the work in this field is focused on identifying situations where we can build such causal models in the absence of any information gained directly via interventions. However, all causal models ultimately gain empirical meaning in virtue of the fact that one can \emph{in principle} test their implications via local interventions. 

In the last few years, physicists have begun to consider the implications this modern approach to causality may have for specifically quantum phenomena. The goals of this recent work are diverse, and the fruits of labour varied, including foundational implications for our understanding of \mbox{causation~\cite{OCB,Cavalcanti, Pienaar, Henson, Spekkens, Chiribella1,Chiribella2, Costa, shrapnel, Allen}}, new insights into non-Markovian quantum dynamics \cite{Pollock, Milz, modi} and practical quantum advantages for certain causal identification tasks \cite{Ried, Chiribella3}. 

A primitive assumption of all these works is the possibility of local interventions \cite{Pearl,Woodward}, controlled operations that can be used to probe and define the causal structure of interacting physical systems. To~date there has been little to no engagement with this aspect of the formalism beyond (i) identifying classical interventions with the setting of a classical random variable to a specific value (so-called ``atomic'' or ``surgical'' interventions)  and (ii) identifying quantum interventions with quantum instruments (formally, completely positive trace preserving maps). It is therefore interesting to ask if there are specific physical constraints that determine whether a given controlled operation can uncover causal structure.

In this paper, we approach this question by considering the abstract notion of an intervention from a physical perspective. Using examples from both classical and quantum mechanics, we ask whether thermodynamic principles place any constraints on the nature of local interventions. 

In Section~\ref{s2}, we summarise the concept of an intervention introduced by Pearl and the causal modelling community. In order to apply this rather abstract definition  to mechanical systems, we define interventions in terms of stochastic control theory based on measurement. This  raises the question of what kinds of interventions are allowed by the laws of physics and the question of an optimal intervention.   In Section \ref{classical}, we show how to describe open systems in classical mechanics in terms of Markov maps. This makes dynamics irreversible and identifies the source of causal asymmetry as unmodelled noise. In Section \ref{interventions},  we show how to define interventions as a control process based on measurement, also described in terms of a Markov map. The intervention cuts the dynamical chain between pre-intervention and post-intervention states, identifying an alternative source of causal asymmetry for causal models.   In Section \ref{thermodynamics}, we  discuss the thermodynamics of interventions as control processes. We define the efficiency of a classical intervention thereby showing that perfect classical interventions are in fact impossible. In Section \ref{quantum}, we turn to the quantum case. We first review the theory of generalised measurements in quantum mechanics and highlight the role of entanglement for quantum open systems. We show, using a simple example of a two-particle collision, how entanglement prevents us from making time reversal transformations on the sub-system. In~Section \ref{quantum-interventions}, we define interventions in the quantum case as a control process described in terms of a completely positive map.  Finally, in Section \ref{q-interventions}, we consider thermodynamic constraints for quantum interventions. Using~the example of coarse-grained position measurements on a thermalised simple harmonic oscillator we show how the uncertainty principle determines the thermodynamic costs of an intervention on a quantum system.  We finish with a discussion.

	\section{Interventions and Causal Processes}\label{s2}

\begin{figure}[H]
\centering
\includegraphics[scale=1.0]{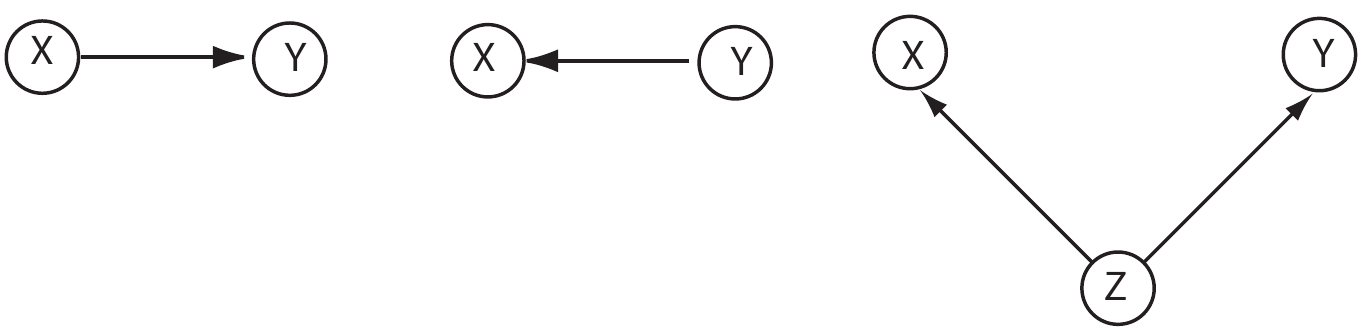}
\caption{Three ways to graphically represent a causal structure that captures the correlation between two variables $X$ and $Y$. }\label{PCC-graph}
\end{figure}

Whilst statistical information can tell us when two variables are correlated, we require explicitly \emph{causal} information to explain how one variable will change in response to changes in another.  Imagine~we make many measurements of two variables, $X$ and $Y$, and see that they are strongly correlated. Without further assumptions, it is impossible to predict what will happen if we intervene and set $X$ to a particular value. One can see that we can explain the correlation between $X$ and $Y$ via three causally distinct hypotheses: $X$ causes $Y$, $Y$ causes $X$, or $X$ and $Y$ are both effects of a common cause $Z$, see~Figure~\ref{PCC-graph}. 

 Without the means to distinguish between these three options, it is impossible to make a predictive statement about what will happen following an intervention that directly sets the value of $X$. What~this means in reality will depend very much on the nature of the variables involved. It is the purpose of this paper to define interventions for mechanical systems and the physical constraints that control~them.
 
 Pearl~\cite{Pearl} defines interventions in a highly abstract way as a `surgery on equations'. Suppose experiments reveal statistical correlations among a set of variables $X_i$ and a putative graphical representation is given for the causal connections between the variables, as in Figure~\ref{PCC-graph}. The  parents of a variable $X_i$ refers to a subset of variables $PA_i$ say$\{X_1,X_2,\ldots X_{i-1}\}$ such that
\begin{equation}
P(x_i|pa_i)=P(x_i|x_1,\ldots,x_{i-1}).
\end{equation}

Each child--parent family is to be represented by a mechanism, or functional relationship,  \mbox{$x_i=f_i(pa_i,u_i)$}, where $u_i$ are independent random disturbances. According to Pearl~\cite{Pearl}:
\begin{quote}
The simplest type of external intervention is one in which a single variable, say $X_i$, is forced to take on some fixed value, $x_i$. Such an intervention, which we call atomic, amounts to lifting $X_i$  from the influence of the old functional mechanism $x_i=f(pa_i,u_i)$, and placing it under the influence of a new mechanism  that sets the value of $x_i$, while keeping all other mechanisms unperturbed.
\end{quote} 

The first question we need to answer is as follows: in a mechanical system of interacting particles, what is meant by `forced'? More significantly, is it possible to satisfy the condition `while keeping all other mechanisms unperturbed' for mechanical systems? In the quantum case, the uncertainty principle may raise some doubts.

The relevant variables of mechanical systems are physical quantities such as position, momentum, energy, etc. We cannot change a physical variable unless we know how much to change it by and that requires making a measurement and then acting, conditionally, with a physically allowed operation on the space of states. Physically allowed means that the mathematical description must correspond to an operation consistent with the laws of physics, including thermodynamics. We thus define interventions in mechanical systems in terms of measurement and control. This necessarily requires us to consider open systems subject to dissipation, noise, and measurement. Causal interventions assume that an agent is acting upon an open subsystem of the world. This helps explain the asymmetrical nature of causal relations in a world in which the global laws of physics are time reversal invariant.

\section{Classical Stochastic Dynamics of Open Systems}\label{classical}

In classical mechanics, the state of a perfectly closed system at any time is given by the position, $ {\bf{q}}(t)=(q_1(t), q_2(t),\dots, q_N(t))$, and momenta, $ {\bf{p}}(t)=(p_1(t),p_2(t),\ldots, p_N(t))$, of all the particles in the system. The dynamics is determined  by a real valued scalar function of all the position and momenta, $H({\bf q,p})$, the Hamiltonian,  using the Poisson bracket operation and Hamilton's equations:
\begin{equation}
\dot{q}_i =  \frac{\partial }{\partial p_i}H({\bf q,p})\ \ \ \ \ ;
\dot{p}_i = -\frac{\partial }{\partial q_i}H({\bf q,p}).
\label{conservative}
\end{equation}

The dynamics is entirely deterministic: Given the Hamiltonian, if we know the position and momenta of every particle at any time $(q_k(t_0), p_k(t_0))$, we can determine the position and momenta of every particle at any point in the future or the past of $t_0$. (Although generic dynamical systems are chaotic and determinism does not necessarily imply predictability.)
  ~As Laplace captured in his famous statement of classical determinism, a complete description of classical dynamics requires global knowledge, in~fact knowledge of the entire universe.  This is never available for any finite agent embedded in the world. How then does one arrive at knowledge regarding  universal laws? It is typically gained in a piecemeal fashion, via local experiment and observation. Causal models are explicitly designed to mirror exactly such local experiments. Such situations are `small worlds', not~only open, but~requiring the inclusion of an experimenter, or control system, to execute the relevant observations and interventions. The key assumption is that such local experiments give us epistemic access to situations where we do not have control, including closed systems, in order to make causal~inferences.

We can illustrate the two complimentary pictures, closed and open, by considering a simple example: the head-on collision of two particles with identical mass. To begin we will assume point particles.  If the degrees of freedom of the two-particle system are isolated from the environment (which could include internal degrees of freedom, here excluded by the assumption of point particles),  Newton's laws of motion tell us that the values of the momenta of each particle are simply exchanged. In this scattering problem, the initial and final momenta are related by $p_{a,f} =   p_{b,i}$ and $p_{b,f}   =    p_{a,i}$.

If one particle is stationary before the collision, it is moving after the collision. It is tempting to say that the moving particle was the \emph{cause}, and the change in the motion of the stationary particle the \emph{effect}. There~is a natural objection to calling this ``causation''~\cite{Russell} as we can swap the labels $i,f$ on each side of the scattering equations and nothing changes. We could just as well say that the final motion is the cause of the earlier motion.  Newton's laws are deterministic and reversible: systems can be perfectly~isolated. 

Even if no external forces act on the two-particle system and total momentum is conserved, there~is the possibility that, if we drop the idealisation of point particles, we need to take into account internal degrees of freedom. These degrees of freedom describe the reversible effects of the internal forces of restitution in the deformation of elastic bodies in collision~\cite{Mach}. The momentum difference of the two particles, $\epsilon=p_{b,i}-p_{a,i}$, becomes correlated with  the  internal degrees of freedom of each particle. Thus,~while the internal degrees of freedom of each particle are uncorrelated (by assumption) before collision, they are necessarily correlated after the collision.  At the collision, the  internal degrees of freedom apply an equal and opposite impulse to each particle: an impulse $\epsilon$ to particle $a$ and an impulse $-\epsilon$ to particle $b$.  This changes the momenta of the particles to $p_{a,f}=p_{b,i}$ and  $p_{b,f}=p_{a,i}$. This description remains time reversal invariant so long as we include both the external and internal degrees of freedom of each particle.

Real experimental systems are, of course, never completely isolated, and measurements are never arbitrarily accurate. This is true of both classical and quantum systems.  This is not just a question of practical considerations.  As we will explain, it is a consequence of the laws of thermodynamics. In~any experiment, a decision must be made as to which degrees of freedom are to be fully accounted for and which are to be treated as unknown, environmental systems. In some cases, as we show, these~unknown degrees of freedom may in fact be internal degrees of freedom that constitute the coarse-grained macroscopic degrees of freedom under experimental control.     

Let us consider the first fact: experimental systems are never isolated.   The standard way to describe this situation is to embed the system of experimental interest into a larger system, called~the environment, and then to engineer the situation so that the energy of interaction between the system and the environment is small on the energy scales relevant for the experimental system itself. As,~by~explicit arrangement, we have little or no knowledge of the microscopic state of the environment, we average over these degrees of freedom to get a statistical (non deterministic) description of the experimental system of interest. In most situations, we assume that the environment is in thermal equilibrium. The net result is that the effect of the environment on the system remains only as a source of dissipation (friction) and small rapidly fluctuating forces.  A controlled mechanical experiment will try to partition the world in such a way that the system-plus-environment approach cuts the world at the relevant degrees of freedom for the experimental investigation by a suitable arrangement of energy scales. It is a remarkable feature of the  physical world that this is possible.

Once we have moved to the system-plus-environment picture, we can no longer specify the state of a physical system as entirely deterministic functions of time, $(q_i(t),p_i(t))$. Instead there are two completely equivalent ways to proceed. We can describe the system dynamics in terms of stochastic functions of time (the `Langevin picture') or we can give a description in terms of a probability density on phase space $\p(q_i,p_i,t)$, such that $\p(q_i,p_i,t)dq_idp_i$ is the probability to find the dynamical state in a small phase space volume centred on $(q_i,p_i)$. We will refer to this as the `Einstein picture', recalling Einstein's approach to diffusion in terms of probability densities. Physical state transformations correspond to Markov maps defined by a Markov kernel as \cite{LasMac,Markov-semigroup}.
\begin{equation}
P(q_i,p_i,t_f)=\int dp_{i}^\prime dq_{i}^\prime K(q_i,p_i, t_f-t_i)|p_{i}^\prime, q^\prime_{i})\p(p_{i}^\prime, q^\prime_{i},t_i).
\end{equation}

The kernel, $K$,  has an interpretation as a conditional probability.  

Using the model with the auxiliary system, we can write 
\begin{equation}
\p_f(p_a,p_b) =\int_{-\infty}^{\infty} d\epsilon  \delta(\epsilon - (p_b-p_a))\p_{a,i}(p_a+\epsilon)\p_{b,i}(p_b-\epsilon).
\end{equation}

The underlying reversible Newtonian  dynamics is reflected in the delta function. This immediately suggests the generalisation.
\begin{equation}
\p_f(p_a,p_b) = \int d\epsilon \p(\epsilon-(p_a-p_b))\p_{a,i}(p_a+\epsilon)\p_{b,i}(p_b-\epsilon)
\end{equation}
where $\p(\epsilon)=P(-\epsilon)$ represents intrinsic initial fluctuations of the internal degrees of freedom.  In this case, the transformation from initial to final distributions is no longer reversible.

To see this, we can easily show that, 
  while mean values are swapped at the collision, \mbox{$E(p_{a,f})=E(p_{b,i})$} and $E(p_{b,f})=E(p_{a,i})$ as we have assumed $\p(\epsilon)=\p(-\epsilon)$, the variances are related by $V(p_{a,f})=V((p_{b,i})+V(\epsilon)$ and $V(p_{b,f})=V((p_{a,i})+V(\epsilon)$. The interaction with unknown internal degrees of freedom always adds noise, which makes the two-particle dynamics irreversible if the auxiliary system is regarded as part of the environment and never observed. While we may continue to regard the final momentum of particle $b$ as the effect that is caused by particle $a$, the process is no longer time reversal invariant.  In this picture, we see that the source of causal asymmetry is simply our ignorance of the unmodelled internal degrees of freedom.

We turn now to consider the concept of an intervention for causal relations in a classical mechanical setting. We show that an intervention is constrained by the thermodynamics of measurement and feedback in open systems.

\section{Classical Interventions}
\label{interventions}

 We take interventions to characterise the act of setting a system to a particular state. This is necessarily a two-stage process: we must first \emph{measure} the system and then implement a controlled operation that is conditioned on the outcome of this measurement. More simply, in order to set a variable to a particular state, one must first know its current state, and   one can then determine the requisite control map. Thus, we will need labels to specify the kind of measurement and the measurement result, as well as labels to specify the type of control map. Intervention thus defined is necessarily asymmetric as the control map must act after the measurement.

The measurement result is generically denoted by $x$ (which could stand for multiple real variables), and   $k$ characterises the kind of measurement {\em and} the nature of the intervention.   A simultaneous measurement of position and momentum as well as a conditional displacement in system momentum might be labelled. Thus, a control map is a Markov map denoted  ${\cal S}_{x_k}$ . That is to say $x_1,x_2, \ldots$ are the measurement results for the corresponding measurement-control maps $k=1,2,\ldots$. Of course, the measurement results need not be canonical variables. For example, we may choose to measure only the kinetic energy of a mechanical system, which is a quadratic function of momentum. However, as all measurements may be regarded as simultaneous measurements of position and momentum followed by some sort of coarse graining, we restrict our attention to such canonical variables.

The measurement is completely characterised by a measurement kernel $M(x|q,p)$, where $x$ are the results of the measurement and $(q,p)$ are the canonical coordinates of the system.  The probability density for the measurement result, and thus for the intervention itself, is given by 
\begin{equation}
\p(x) =\int dqdp M(x|q,p)\p^s(q,p)
\end{equation}
where $\p^s(q,p)$ is the state of the system prior to the intervention. 

The conditional state of the system, given the measurement result, $x$, is 
\begin{equation}
\p^s(q,p|x) =\frac{M(x|q,p)\p^s(q,p)}{\p(x)}.
\end{equation}

We define a conditional control map by the Markov map with the stochastic kernel $C(q,p,|q',p',x)$. The {\em conditional} system state, after the intervention, is given by 
\begin{equation}
 \p_c^s(q,p|x)=\left (\p(x)\right )^{-1}\int dq'dp'\ C(q,p|q',p', x)M(x|q',p')\p^s(q',p')
 \end{equation}

Thus, the intervention is completely characterised by the  stochastic kernels $C(q,p|q',p', x)$ and the measurement kernel $M(x|q',p')$ corresponding to the control map and the measurement map respectively. The probability distribution for the intervention $\p(x)$ is given by 
 \begin{eqnarray}
 \label{cl-intervention-prob}
 \p(x) & = &  \int dqdp dq'dp' C(q,p|q',p', x)M(x|q',p') \p^s(q',p')\\
 & = & \int dqdp dq'dp'   S(q,p|q',p',x)   \p^s(q',p')\\
 & = & || {\cal S}_x \p^s||
 \end{eqnarray}
 with the $L^1$ norm on phase space. 

Finally, to complete the description, we can compute the {\em unconditional} system state after the intervention:
 \begin{equation}
 \bar{\p}^s(q,p)=\int dxdq'dp'\ C(q,p|q',p' ,x)M(x|q',p')\p^s(q',p').
 \end{equation}
 
In general, the intervention takes many system states to one particular unconditional system state. If we keep a record of both the measurement result and the control map applied, then the intervention is reversible. Typically, however, interventions are labelled only by the final desired state. In this sense, the intervention is irreversible: we cannot determine what state we started with after the intervention has taken place if we discard the measurement result.  The intervention removes correlations between the system after the intervention and earlier states. This arises due to the act of measurement itself which requires that the system be strongly coupled to a low entropy apparatus. Our ignorance of the pre-intervention state of the \emph{system} is itself due to unmodelled features of the environment (internal or external) as suggested in Section~\ref{classical}.

\section{Thermodynamics of Interventions in Classical Mechanics}\label{thermodynamics} 
We take the view that a control map, taking a conditional state resulting from a measurement to the desired state after the intervention, must also be implemented in a completely mechanical fashion by acting on the system with mechanical forces. Given this constraint, it is clear that an intervention can be  regarded as a problem in classical stochastic control. Once this is realised, the question of optimality can be considered. 

Two distinct intervention maps ${\cal S}_x$ and ${\cal R}_x$ such that 
\begin{equation}
\int dq' dp'S(q,p|q',p',x)\p^s(q',p') = \int dq' dp'R(q,p|q',p',x)\p^s(q',p')
\end{equation}
are equivalent. Note that, if two interventions are equivalent, then
\begin{equation}
||{\cal S}_x\p^s||= ||{\cal R}_x\p^s|| = \p(x).
\end{equation}

Thus, if two interventions are equivalent, the corresponding conditional states they result in are~identical. 

Is there any way to choose between equivalent interventions? In real physical systems, some may be more difficult to implement than others. Indeed, some may be impossible. How can we quantify the difficulty (or even the feasibility) of achieving a given intervention?  We shall approach this question by considering if there are any  important thermodynamic constraints on possible interventions.

As an example, consider interventions on a single free particle in thermal equilibrium. The~mathematical details of this example are given in Appendix~\ref{SA1},  where we give an explicit model of a measurement on a free particle in terms of its interaction with another physical system, the~apparatus. The state of the apparatus is assumed to be entirely under our control, while the state of the system is unknown.  Here we summarise those results.  The intervention objective is to control the momentum so that it is highly localised on $p=0$. We will assume that initially the particle is in thermal  equilibrium at temperature $T$ so that the state is given by a Gaussian distribution in momentum with mean zero and variance $\Delta=k_BT$. Suppose now we make a measurement of momentum, with a Gaussian measurement kernel with momentum variance $\sigma$. As shown in Appendix~\ref{SA1}, this uncertainty is determined by the uncertainty in the state of the apparatus. Let the result of the measurement be $x$. The conditional state of the system, given $x$, is a Gaussian with mean $E(p|x)=x(C-1)/C$ and variance $V(p|x)=\Delta/C$, where $C=1+\Delta/\sigma$. For a very accurate measurement, $\sigma <<\Delta $ so that $C >>1$. In fact, with the limit of no added measurement noise, $\sigma\rightarrow 0$ and $C\rightarrow \infty$; the uncertainty in the system momentum in the conditional state vanishes, resulting in a perfectly accurate measurement. In~other words, in order to make the measurement, an agent must prepare the entropy of the apparatus to be much less than the entropy of the system to be measured (Appendix~\ref{SA1}).  This asymmetry is required for an accurate measurement.  

A simple intervention protocol is obvious: subject the particle to an impulsive force to shift the mean momentum by $p_0=-x(C-1)/C$.  The final state is a Gaussian, with mean momentum zero and a variance of $\Delta/C$. The average energy of the displaced state is due entirely to the momentum fluctuations and is  $\Delta/(2mC)$. 

The intervention comes with an energy cost. The work done by the control protocol is $p_0^2/2m$.  This is a random variable as the result of the measurement is a random variable. The average work done {\em on} the system over a large number of trials is  
\begin{equation}
\bar{W}= \frac{\Delta}{2m}\left (\frac{C}{C-1}\right ),
\end{equation}
and the change of the average energy of the system, after the intervention, is 
\begin{equation}
\Delta\bar{U} =  -\frac{\Delta}{2m}\left (\frac{C-1}{C}\right ).
\end{equation}

As $C= 1+\Delta/\sigma$, this is negative. In the case of a perfect intervention, $C\rightarrow \infty$ and $\Delta\bar{U}=-\bar{W}$.   

We define the efficiency of the intervention as
\begin{equation}
\eta =\frac{|\Delta\bar{U}|}{\bar{W}}.
\end{equation}

In the Gaussian model we are considering, the efficiency is given by 
\begin{equation}
\eta=\left (\frac{C-1}{C}\right )^2,
\end{equation}
which is always less than unity and tends to unity in the limit of $C\rightarrow \infty $ for perfect interventions. 

 In general, this is an upper bound as the control process that shifts the momentum may add noise unlike the simple conservative displacement we have assumed above.  This could occur if the memory that stores the measurement result is subject to errors through its interaction with an unknown environment.  If the shift increases the variance, along with a systematic reduction in the kinetic energy, the intervention effectively heats the system and does 
  work on it.  

In the previous model, a perfect intervention is only possible if the entropy of the apparatus is zero, $\sigma\rightarrow 0$, prior to measurement.  This means that the measurement apparatus (and its memory which stores the measurement to enable an intervention) must have had zero entropy prior to the intervention~\cite{Jacobs}. In other words, the measurement apparatus must have been cooled using a zero temperature heat bath.  In reality we do not have access to a zero temperature heat bath, and we must prepare the state of the measurement apparatus (including the memory) by a finite process that reduces its entropy as best as we can.  The measurement result has a finite probability of being in error even if the interaction between the system and apparatus is perfectly reversible. Interventions are necessarily irreversible process as far as the sub-system intervened upon is concerned. This kind of causal asymmetry, the asymmetry introduced by interventions, is underwritten by the third law of thermodynamics and clearly physically unavoidable.

 The change in entropy between the final and initial system distributions is 
\begin{equation}
\Delta S=-\frac{1}{2}\ln(C)\ ,
\end{equation}
which is negative. On the other hand, the average mutual information is
\begin{equation}
\bar{I}= \frac{1}{2}\ln(C)\ .
\end{equation}

We can put these results into a more general context for a classical feedback processes by using the stochastic thermodynamics with feedback considered by Sagawa and Ueda~\cite{sagawa1,sagawa2}. After the measurement, we let the system return to thermal equilibrium at the initial temperature.  In this example, we find that the change in free energy of the system 
\begin{equation}
\Delta F = -\frac{\Delta}{2m}\left (\frac{C-1}{C}\right )+\frac{k_BT_s}{2}\ln(C).
\end{equation}

During this process, we can extract work $\bar{W}_{ext}$ such that
\begin{equation}
\bar{W}_{ext}\leq -\Delta F+k_BT_s\bar{I}=\frac{\Delta}{2m}\left (\frac{C-1}{C}\right ).
\end{equation}

If we compare this to the average work done on the system by the intervention, we see that, in general, $\bar{W}_{ext} \leq \bar{W}$ with equality only in the case of a perfect intervention $C\rightarrow \infty$. 
 
Let us return to the question of an optimal intervention starting with an initial thermal momentum distribution. These two interventions are equivalent:
\begin{itemize}
\item A Gaussian measurement that reduces the momentum variance to $\Delta$ followed by a conservative shift to average zero momentum.
\item A more accurate Gaussian measurement that reduces the momentum variance to $\Delta-\mu$ followed by a noisy shift
that increases 
 the variance to $\Delta$ and shifts the average momentum to zero.  
\end{itemize}

The corresponding intervention kernels are 
\begin{eqnarray}
S_1(q,p|q',p',p_0) & = &  \delta(p'-p-p_0) M_1(p_0|p') \\
S_2(q,p|q',p',p_0)  & = & (2\pi\mu)^{-1/2}e^{-(p'-p-p_0)^2/2\mu}M_2(p_0|p')
\end{eqnarray}
where the measurement kernels are defined by
\begin{equation}
M_k(p_0|p) = (2\pi\sigma_k)^{-1/2}e^{-p^2/2\sigma_k}
\end{equation}
where 
\begin{eqnarray}
\sigma_1 & = & \left [\frac{1}{\Delta}-\frac{1}{mk_BT}\right ]^{-1} \\
\sigma_2 & = & \left [\frac{1}{\Delta-\mu}-\frac{1}{mk_BT}\right ]^{-1}.
\end{eqnarray}

However, the second intervention requires more energy on average by the intervention control process, so the
corresponding intervention map is less efficient than the first. See Appendix~\ref{SA2}. 

\section{Quantum Open Systems}\label{quantum}
We now consider causal interventions in the quantum case. As with the classical case, we~first consider the source of causal asymmetry in the absence of interventions. Let us return to the simple collision model in a quantum description. There are two equivalent ways to proceed: via the  Heisenberg picture or via the Schr\"{o}dinger picture. This mirrors the Langevin versus the Einstein picture in the classical case.  We will begin with the Schr\"{o}dinger picture.

The quantum scattering theory for a two particle collision of this kind is known, but we do not need it here. A simple model that captures the key features of the classical collision can be defined using a quantum variant of the classical model with elastic internal degrees of freedom, called the auxiliary system. The initial state is taken as 
\begin{equation}
|\Psi_i\rangle  =|\psi\rangle_a\otimes|\phi\rangle_b\otimes|0\rangle_c
\end{equation}
where $|\psi\rangle_a$ and $|\phi\rangle_b$ are arbitrary single particle states, and $|0\rangle_c$ is a fiducial initial state for the auxiliary system.
We define the interaction between the particles and the auxiliary system by the pair of unitaries
\begin{eqnarray}
U & = & \exp[\frac{i}{\hbar}\hat{Q} (\hat{p}_b-\hat{p}_a)]\\
V & = & \exp[-\frac{i}{\hbar}\hat{P} (\hat{q}_b-\hat{q}_a)]
\end{eqnarray}
where $\hat{Q},\hat{P}$ are operators on the auxiliary system such that $[\hat{Q},\hat{P}]=i\hbar$. We will now define the fiducial state $|0\rangle_c$ in the diagonal basis of $\hat{P}$:
\begin{equation}
|0\rangle_c =\int_{-\infty}^{\infty} d\epsilon {\cal A}(\epsilon)|\epsilon\rangle_c
\end{equation}
where  $\hat{P}|\epsilon\rangle_c =\epsilon|\epsilon\rangle_c$.

The total state after the interaction is defined by $|\Psi\rangle_f= VU|\Psi\rangle_i$ and can be shown to be 
\begin{equation}
|\Psi_f\rangle = \int_{-\infty}^{\infty} d\epsilon\    dp dp'  {\cal A }(\epsilon-(p'-p))\psi(p+\epsilon)\phi(p'-\epsilon)|p\rangle_a\otimes|p'\rangle_b\otimes|\epsilon\rangle_c,
\end{equation}
assuming ${\cal A}(\epsilon)= {\cal A}(-\epsilon)$. 
It is easily seen that this model conserves total momentum of particle $a$ and particle $b$. 

A special case occurs for ${\cal A }(\epsilon-(p'-p))=\delta(\epsilon-(p'-p))$. Integrating over $\epsilon$, we see that the state of the two particles factors out as 
\begin{equation}
|\psi_f\rangle_{ab} = |\phi\rangle_a\otimes|\psi\rangle_b.
\end{equation}

This is a perfect state swap and of course it is unphysical as it requires preparing the auxiliary system in an eigenstate of $\hat{P}$, which is physically impossible; a consequence of the uncertainty principle for $\hat{Q}$ and $\hat{P}$.  In the physically realistic case, the two particles remain entangled with the auxiliary system after the interaction. The degree of entanglement depends on the extent of delocalisation of the initial wave-function ${\cal A}(\epsilon)$, equivalently, how well defined the variable $\hat{Q}$ is in the initial state. The~fact that the final total physical state is bi-partite entangled  (two-particle system $\times$ internal system) implies that we cannot reverse time on the two-particle subsystem alone and maintain positivity of the total state, as this is equivalent to partial transposition on the two-particle subsystem~\cite{Simon}. Only global time reversal is physically permitted.   This would suggest, when we limit our focus to the two particle sub-system, it is not only our ignorance of the state of the unmodelled internal degrees of freedom but~also  the uncertainty principle that serves as the source of causal asymmetry in the quantum case. 

We can calculate the reduced state of the two-particle system by tracing out the auxiliary system.  This is given by
\begin{equation}
\label{unconditional-collision}
\rho_{ab,f}=\int_{-\infty}^\infty\ d\epsilon\  e^{-i\epsilon(\hat{q}_b-\hat{q}_a)/\hbar}\hat{{\cal A}}(\epsilon-(\hat{p}_b-\hat{p}_a))\ \rho_{ab,i}\ \hat{{\cal A}}^\dagger(\epsilon-(\hat{p}_b-\hat{p}_a))e^{i\epsilon(\hat{q}_b-\hat{q}_a)/\hbar}
\end{equation}
where in this case the initial pure state is $\rho_{ab,i}= |\psi\rangle_a\langle\psi|\otimes|\phi\rangle_b\langle\phi|$. This has a direct interpretation as a measurement and control operation on the two-particle system:  the internal degree of freedom is first coupled via $\hat{U}$,   $\hat{P}$ is measured, and the momentum difference between the two particles is then shifted 
  using $\hat{V}$ by the measured result.

We can also interpret this model in the Heisenberg picture in which states are unchanged and operators are transformed by $U^\dagger V^\dagger \hat{A}  VU$. For example,  $\hat{p}_{a,f}= \hat{p}_{b,i}+\hat{P}_i$. If we choose the state of the auxiliary system so that $\langle \hat{P}_i\rangle =0$, then, on average, $\langle \hat{p}_{a,f}\rangle = \langle \hat{p}_{b,i} \rangle$, as we expect from the classical analogy. On the other hand, the auxiliary system always adds noise as we see by calculating the~variances
\begin{equation}
\langle \Delta \hat{p}_{a,f}^2\rangle = \langle \Delta \hat{p}_{b,i}^2\rangle +\langle \Delta\hat{P}_i^2\rangle.
\end{equation}

Only in the unphysical case of the auxiliary system prepared in an eigenstate of $\hat{P}_i$ are the states completely swapped between input and output.

In the quantum case, the probability distribution for measurement results $x\in \Re$  are given by 
\begin{equation}
P(x) = {\rm tr}[ \rho\hat{E}(x)]
\end{equation}
with $\rho$ the density operator representing the state of the particle and $\hat{E}(x)$ is a positive operator such that $\int_{-\infty}^\infty dx \hat{E}(x) =1$. We can thus write 
\begin{equation}
 \hat{E}(x)= \hat{\Upsilon}^\dagger (x)\hat{\Upsilon}(x).
\end{equation}

The conditional state of the particle after measurement, given the result $x$, is  
\begin{equation}
\label{q-meaasurement-conditional}
\rho_{|x }= \frac{\hat{\Upsilon}(x) \rho \hat{\Upsilon}^\dagger (x)}{P(x)},
\end{equation}
while the unconditional state of the measured particle is given by 
\begin{equation}
\rho' = \int_{-\infty}^{\infty} dx \hat{\Upsilon}(x) \rho \hat{\Upsilon}^\dagger (x).
\end{equation}

An example for the case of position measurements is given in~\cite{CavesMil87}. 

\section{Quantum Interventions}
\label{quantum-interventions}
The simplest intervention is a quantum version of the measure and control 
protocol of the classical case. Given the measurement result $x$, we apply a conditional unitary to the measured system. Thus,~the system 
state after the control intervention is given by 
\begin{equation}
\rho_{|x}= \frac{U(x)\hat{\Upsilon}(x) \rho \hat{\Upsilon}^\dagger (x)U^\dagger(x)}{P(x)}.
\end{equation}

The corresponding unconditional state is
\begin{equation}
\rho'= \int dx U(x)\hat{\Upsilon}(x) \rho \hat{\Upsilon}^\dagger (x)U^\dagger(x).
\end{equation}

Note that the order of the conditional unitary and the measurement operator is important as they may not commute and it is essential that the feedback control term acts after the measurement result. An example of this is the two-particle model described in the previous section, as is evident from Equation (\ref{unconditional-collision}).  In a more general setting, the control transformations are not necessarily unitaries but any CPTP map so that
\begin{equation}
\rho_{|x}= {\cal E}(x)\left [\frac{\hat{\Upsilon}(x) \rho \hat{\Upsilon}^\dagger (x) }{P(x)}\right ].\ 
\end{equation}

 For example, after the measurement, the system may be permitted to come into thermal equilibrium with a reservoir at a temperature that depends on $x$.

In analogy with the classical case, we will define an intervention map by the completely positive one-parameter map:
\begin{equation}
{\cal S}(x)\rho ={\cal E}(x)\left (\hat{\Upsilon}(x) \rho \hat{\Upsilon}^\dagger(x)\right )
\end{equation}
where $x$ is a real valued random variable  and the corresponding probability of this particular intervention~is 
\begin{equation}
\p(x) = {\rm tr}[{\cal S}(x)\rho ].
\end{equation}

This is the quantum analogue of Equation (\ref{cl-intervention-prob}).

\section{Thermodynamics of Quantum Interventions}
\label{q-interventions}
A quantum intervention as we have  defined it is constrained by the thermodynamics of measurement and feedback~\cite{sagawa1,sagawa2, Rudolph}. As an example, we will consider a binary measurement on a simple harmonic oscillator.  

First, consider a measurement that simply asks  if a particle moving in a quadratic potential, centred on the origin, is   on the left of the origin or on the right? This kind of measurement is used in the Szilard engine model for a gas in a cylinder.  

Such a measurement can be described by the measurement operators~\cite{khosla}:
\begin{equation}
\hat{M}_{\pm}=\frac{1}{2}\left [1\mp i\frac{(\lambda+i\hat{q})}{\sqrt{\lambda^2+\hat{q}^2}}\right ]
\end{equation}
where the parameter $\lambda$ controls how closely this corresponds to a measurement of the sign of the displacement (see below).
The probabilities of the measurement results are determined by 
\begin{equation}
\p(\pm)={\rm Tr}[\hat{M}^\dagger_{\pm}\hat{M}_{\pm}\rho]
\end{equation}
where 
\begin{equation}
\hat{M}^\dagger_{\pm}\hat{M}_{\pm}=\frac{1}{2}\left [1\pm \frac{\hat{q}}{\sqrt{\lambda^2+\hat{q}^2}}\right ].
\end{equation}

In the limit $\lambda\rightarrow 0$, this approaches a measurement of the sign of the displacement from equilibrium. Clearly $\hat{M}^\dagger_{+}\hat{M}_{+}+\hat{M}^\dagger_{-}\hat{M}_{-}=1$. 

It is easy to see that, when acting on energy eigenstates,
\begin{equation}
\hat{M}_{\pm}|n\rangle=\frac{1}{2}(|n\rangle \pm |\psi_n\rangle) \equiv \frac{1}{\sqrt{2}}|\phi_n^{\pm}\rangle 
\end{equation}
where $|\psi_n\rangle$ is a parity eigenstate with eigenvalue $(-1)^{n+1}$; that is to say, it has the opposite parity to $|n\rangle$, and the unitary parity operator is defined $\hat{\Pi}=e^{-i\pi a^\dagger a}$, with $a,a^\dagger$ the usual raising and lowering operators for the oscillator. Thus, $|n\rangle$ and $|\psi_n\rangle$ are orthogonal for every $n$. The states, $|\phi_n^{\pm}\rangle$, are not orthogonal, as
\begin{equation}
\langle \phi_m^+| \phi_n^+\rangle= \delta_{m,n}+g_n(\lambda)(\delta_{m,n+1}+\delta_{m,n-1})
\end{equation}
where $g_n(\lambda)=\langle n|[\mu^2+(a+a^\dagger)^2]^{-1/2}|n\rangle$ and 
$\mu=\lambda/\sqrt{\Delta_0}$, and $\Delta_0$ constitutes the rms fluctuations of position in the oscillator ground state.

The average displacement in the conditional states, $|\phi_n^{\pm}\rangle$, is given by 
\begin{equation}
\langle \phi_n^{+}|\hat{q}|\phi_n^{+}\rangle =-\langle \phi_n^{-}|\hat{q}|\phi_n^{-}\rangle=\frac{1}{2}(\langle n|\hat{q}|\psi_n\rangle+\langle \psi_n|\hat{q}|n\rangle).
\end{equation}

The conditional states are located on opposite sides of the origin as expected. The average energy in the conditional states is
\begin{equation}
\hbar\omega \langle \phi_n^{\pm}|a^\dagger a|\phi_n^{\pm}\rangle=\frac{\hbar\omega}{2}(n+\langle \psi_n|a^\dagger a|\psi_n\rangle).
\end{equation}

Thus, the measurement on average adds energy, $\Delta E_n=\langle \psi_n|a^\dagger a|\psi_n\rangle$, to the measured system. This~is a direct consequence of the uncertainty principle as the conditional states have reduced uncertainty in displacement. This energy is supplied by the measurement apparatus itself or, more~precisely, by~the classical field that controls the coupling of the apparatus to the system.

If the harmonic oscillator is initially in a thermal state at temperature $T$, the unconditional state after the measurement is 
\begin{equation}
\rho'=\frac{1}{2}\sum_{n=0}^\infty p_n [|\phi_n^+\rangle \langle \phi_n^+|+|\phi_n^-\rangle \langle \phi_n^-|],
\end{equation}
with $p_n=(1+\bar{n})^{-1} [\bar{n}/(1+\bar{n})]^n$ and the average excitation number for an oscillator in thermal equilibrium with a bath at temperature $T$ is $\bar{n} =(\exp(\beta \hbar\omega )-1)^{-1}$, where $\beta^{-1}=k_BT$. This is an equal mixture of two states which each separately have the same entropy equal to the thermal Shannon entropy $H({p_n})$. As one bit is required to specify the components of the mixture, the total entropy of the unconditional state is $S(\rho')=H({p_n})+k_B\ln 2$. The extra bit of information is the mutual information between the system and the memory.

As an example of an intervention we will use the unity parity operator as the feedback operator: if the result of the measurement is $+$, do nothing; else apply the parity operator. The state after the intervention is thus 
\begin{equation}
\rho_f= \sum_{n=0}^\infty p_n |\phi_n^+\rangle \langle \phi_n^+|,
\end{equation}
which has one bit of entropy less than the post-measurement unconditional state, while one bit of entropy is stored in the control apparatus. As the parity operator simply represents free evolution of the oscillator over one period, the feedback does not change the average energy of the state. Thus, after the intervention the change in the average energy of the state due to the intervention is simply
\begin{equation}
\overline {\Delta E}= \sum_{n=0}^\infty \Delta E_n
\end{equation}
where $\Delta E_n=\langle \psi_n|a^\dagger a|\psi_n\rangle$. 

This additional energy can be estimated using the uncertainty principle. The average energy of a simple harmonic oscillator is bounded by the variance of the canonical variables,
\begin{equation}
\bar{E} \leq \frac{V(\hat{p})}{2m}+\frac{m\omega^2}{2} V(\hat{q})
\end{equation}
where $m$ is the mass of the particle and $\omega$ the frequency of oscillator. For a thermal state, this becomes an equality
\begin{equation}
\bar{E}=(2\bar{n}+1)\hbar\omega/2
\end{equation}
where $\bar{n}$ is the average thermal population of the oscillator energy eigenstates. A measurement of which side of the origin the particle is on will reduce the variance in position by a factor of $\gamma<1$. The~variance in momentum will necessarily increase by at least a factor of $1/\gamma$. Thus, the energy bound increases to
\begin{equation}
\bar{E}\leq(2\bar{n}+1)\frac{\hbar\omega}{2}(\gamma+\frac{1}{\gamma}).
\end{equation}

Supposing that $\gamma <1/2$, we see that, after the intervention, the average energy increase of the oscillator~is 
\begin{equation}
\overline{\Delta E}\leq(2\bar{n}+1)\frac{\hbar\omega}{4}.
\end{equation}

At high temperatures, this becomes $\overline{\Delta E}\leq k_BT/2\leq k_BT \ln 2$.

We now ask for the change in free energy when the state $\rho_f$ is allowed to return to thermal equilibrium by interacting with a heat-bath at temperature $T$. This is described by the CP map
\begin{equation}
{\cal E}(\rho)= \sum_{n=0}^\infty \sigma_n^\dagger \rho \sigma_n
\end{equation}
where the set of lowering operators is defined, for each $n$, by $\sigma _n=|n\rangle\langle \psi_n|$. The entropy is unchanged and the energy of the system changes by $-\Delta E$. Thus, the change in free energy of the system is $\Delta F_s = -\Delta E$. The maximum amount of work, $W_{ext}$, that can be extracted during this thermalisation is constrained by  $W_{ext}\leq \Delta E$. As in the classical case, the amount of work that can be extracted is less than the amount of energy supplied by the intervention. However, in this case, this energy is required by the uncertainty principle that links the accuracy of a coarse-grained measurement of position of the system to the uncertainty in momentum of the system. At high temperatures, the uncertainty principle argument of the preceding paragraph shows that 
$W_{ext}\leq k_BT\ln 2$. That is to say, the maximum amount of work that can be extracted by letting the system, post-intervention, return to thermal equilibrium is bounded by the erasure cost of the memory of the intervention. We could simply have extracted work by letting the memory return to thermal equilibrium instead.  

\section{Conclusions}

We have defined an intervention for both classical and quantum systems in terms of a stochastic control operation. Necessarily this requires that an agent partitions the world into system of interest and everything else (internal or external) which is to be treated as an environment.  If such interventions are used to discover causal relations,  the causal data in both the classical and quantum context have the same status: they are experimental measurement data. In both the classical and quantum case, we have argued that no measurement can be perfectly accurate as no laboratory system can ever be completely isolated from its environment. Using simple, but physically reasonable examples, we have illustrated that the known thermodynamics of control operations constrains possible interventions in both the classical and the quantum case. In particular, we have shown how unique quantum features such as entanglement and the uncertainty principle ensure that these constraints are satisfied in the quantum case.  

Our work highlights a possible answer to an interesting question: is the source of causal asymmetry between cause and effect simply the act of intervention itself? Intervention based on measurement and control requires a local low-entropy agent.  The reversibility of dynamical laws suggests the possibility that agent-based interventions are required to endow a causal model with directionality. Indeed, many philosophers would consider that causation should be regarded as an explicitly agent-dependent notion: without interventions, causality simply disappears, leaving only the temporally symmetric dynamical laws in play. An alternative perspective, however, states that physical systems can possess causal properties in the absence of the interventions used to reveal them~\cite{Pearl}. Our view is somewhere between these two extremes: it is the temporally symmetric laws of physics that underwrite the agent-based interventions through which asymmetric causal relations are~discovered.  

While we have focussed exclusively on causal interventions, our approach indicates a feature of causal relations in general. The description of general measurements suggests that the relation between system and apparatus is analogous to that between cause and effect. Given that all physical interactions are reciprocal, what distinguishes the system and apparatus is the lower entropy of the apparatus.  For an accurate measurement to be possible, the \emph{a priori} entropy of the apparatus must be less than that of the system. This suggests that causal relations built on reversible physical interactions require  a local entropy gradient between cause and effect prior to interventions.  This latter conjecture lends support to the hypothesis that causal structure is an agent-independent but contingent property of the world, and interventions merely provide the epistemic route to causal discovery. This challenges accounts of causation that prefer to reduce causation to agency.
\vspace{6pt}



\subsection*{Acknowledgments}
This research was supported by the Australian Research Council Centre of Excellence for Engineered Quantum Systems (project ID CE170100009).

\section*{Appendix A}
In this appendix, we give an explicit example of a measurement based intervention for Gaussian states. In the case of one degree of freedom, these take the form 
\begin{equation}
\p(q,p) =\left (4\pi^2\det\Delta\right )^{-1/2}\exp\left [-\frac{1}{2}(\vec{u}-\vec{u}_0)\Delta^{-1}(\vec{u}-\vec{u}_0)^T\right ]
\end{equation}
where we have defined the vector $\vec{u}= (q,p)$ and $\vec{u}_0=(E(q), E(p))$,
while the covariance matrix, $\Delta$,~is~defined by $
 \Delta_{ij} =E(u_iu_j)-E(u_i)E(u_j)
$ for simplicity let us assume that there are no correlations between position and momentum $E(qp)-E(q)E(p)=0$. 

For the measurement model discussed in Section \ref{interventions}, we only require the marginal distributions for the momentum canonical variable. We thus take the system state before the measurement to be 
\begin{equation}
\p^s_i(p) =\int dq P^s_i(q,p) dp = \left (2\pi \Delta\right )^{-1/2} \exp\left [-\frac{(p-p_0)^2}{2\Delta}\right ],
\end{equation}
while for the apparatus we take 
\begin{equation}
\p^i_a(x) = \left (2\pi \sigma\right )^{-1/2} \exp\left [-\frac{x^2}{2\sigma}\right ].
\end{equation}

The total system-apparatus marginal distribution after the measurement can be defined by a shift map on the apparatus as
\begin{equation}
\p^{s,a}= \p^i_a(x-\mu p)\p^s_i(p).
\end{equation}

This describes a conservative interaction between system and apparatus. Thus, the measurement kernel is $M(p|x)=\p^i_a(x-\mu p)$.

The reduced state of the apparatus after the interaction with the measurement apparatus is then given by a Gaussian with mean
\begin{equation}
E(x) =\mu p_0
\end{equation}
and variance 
\begin{equation}
V(x) \equiv E(x^2)-E(x)^2=\sigma+\mu^2 \Delta.
\end{equation}

Thus, the mean of this distribution is proportional to the mean of the prior distribution for the system momentum, while the variance is the sum of the variance in the apparatus state  and a rescaled variance of the prior system momentum variance. If the prior state of the meter is well defined  so that $\sigma\rightarrow 0$, the measurement statistics are the same as the prior momentum distribution of the system with a rescaling by $\mu$. We can thus regard $\sigma$ as the noise added by the measurement apparatus. This is an important point: to effect a good measurement, the entropy of the apparatus must be much less than the entropy of the measured system.  This asymmetry is explicitly introduced by whatever agent is preparing the apparatus prior to making the measurement.   

The marginal distribution of the system momentum for the {\em conditional} state of the system, given a measurement record $x$, is   also a Gaussian with mean and variance given
\begin{eqnarray}
E(p|x) & = &p_0+\frac{C-1}{C}(x/\mu-p_0) \\
V(p|x) & = & \Delta/C
 \end{eqnarray}
 where 
 \begin{equation}
 C= 1+\frac{\mu^2\Delta}{\sigma}.
 \end{equation}
 
Note that the variance changes in a deterministic way while the change in the average is a random variable, as $x$, the measurement result, is a random variable. As $C>1$, we see that the variance of the momentum in the conditional system state of the system is always reduced.  In the limit of no added measurement noise, $\sigma\rightarrow 0$ and $C\rightarrow \infty$; the uncertainty in the system momentum in the conditional state vanishes: a perfectly accurate measurement.  

The average mutual information between the apparatus and the system is defined by 
\begin{equation}
\bar{I}= \int dp \p(p)\int dx \p(x|p)\ln \left (\frac{\p(x|p)}{\p(p)}\right ).
\end{equation}

Using 
\begin{eqnarray}
\p(x|p) & = & (2\pi\sigma)^{-1/2} e^{-(x-p)^2/2\sigma}\\
\p(x) & = & (2\pi(\Delta+\sigma))^{-1/2} e^{-x^2/(2(\sigma+\Delta))}\\
\p(p) & = &  (2\pi\Delta)^{-1/2}e^{-p^2/2\Delta},
\end{eqnarray}
we find that $\bar{I}= \frac{1}{2}\ln C$.

\section*{Appendix B}
We define a noisy shift in momentum is defined by a convolution kernel.  Let the initial distribution be a Gaussian
conditionally centred on the measurement result $p_0$:
\begin{equation}
\p_i(p|p_0)= (2\pi\Delta)^{-1/2}e^{-(p-p_0)^2/2\Delta}.
\end{equation}

After the noisy shift, the final distribution is 
\begin{equation}
\p_f(p) = \int_{-\infty}^{\infty} (2\pi\mu)^{-1/2} e^{(p'-p-p_0)^2/2\mu}\p_i(p'|p_0).
\end{equation}

This is a Gaussian with zero mean and variance $\Delta+\mu$. 

In this case, the intervention must not only change the average momentum of the particle but must `heat' it so as to increase the 
variance in momentum. Thus the change in energy of the system  is 
\begin{equation}
\overline{\Delta E}=\frac{p_0^2}{2m}+\frac{\mu}{2m}.
\end{equation}


\end{document}